\documentclass[pre,superscriptaddress,twocolumn,a4paper,showpacs]{revtex4-1}
\usepackage{graphics}
\usepackage{amssymb}
\usepackage{wasysym}
\usepackage{latexsym}
\usepackage{graphicx}
\usepackage{epsfig}
\usepackage{subfigure}
\usepackage{dcolumn}
\usepackage{bm}
\usepackage{comment}
\usepackage{tabularx}
\usepackage{adjustbox}
\usepackage{color, colortbl}
\definecolor{Gray}{gray}{0.9}
\begin{document}

\title{A SIR epidemic model for citation dynamics }

\author{Sandro M. Reia}
\affiliation{Instituto de F\'{\i}sica de S\~ao Carlos,
  Universidade de S\~ao Paulo,
  Caixa Postal 369, 13560-970 S\~ao Carlos, S\~ao Paulo, Brazil} 
  
\author{Jos\'e F. Fontanari}
\affiliation{Instituto de F\'{\i}sica de S\~ao Carlos,
  Universidade de S\~ao Paulo,
  Caixa Postal 369, 13560-970 S\~ao Carlos, S\~ao Paulo, Brazil}


\begin{abstract}
The study of citations in the scientific literature  crosses the boundaries between the traditional branches of science and stands on its  own as a  most profitable research field  dubbed  the  `science of science'.  Although  the understanding of the citation histories of individual papers involves  many intangible factors, the basic assumption that citations beget citations can explain most  features of the empirical citation patterns. Here we use the SIR epidemic model as  a mechanistic model for the citation dynamics of well-cited papers  published in selected  journals of the American Physical Society. The estimated  epidemiological parameters offer insight on unknown quantities as the size of the community that could cite a paper and  its  ultimate impact on that community. We find a good, though imperfect,  agreement between the rank of the  journals obtained  using the epidemiological parameters and the impact factor rank.

\end{abstract}

\maketitle

\section{Introduction}

Regardless  of the controversial and widespread use of citation based  measures as a  quantitative proxy of a paper's importance \cite{Merton_1973,Price_1975,Radicchi_2008,Radicchi_2009,Uzzi_2013,Ioannidis_2014},   the study of citations seems to have acquired a life of its own \cite{Mingers_2015}.  In fact,  citation networks,  citation distributions and citation dynamics are topics that range over  most of the issues addressed by the  science of complexity. In addition,  the large citation datasets, which unfortunately are rarely freely accessible,  makes the subject very 
attractive since, contrary to  most complex systems problems, the theories about citation patterns can  readily be  tested against empirical data \cite{Redner_1998,Redner_2005,Wallace_2009}.

A remarkable outcome of the quantitative study of the   citation patterns is the realization that starkly different citation histories,  such as 
the rare `sleeping beauties' (i.e., papers that are not cited  for a long while and then suddenly become popular \cite{van2004sleeping,ke2015defining}) or the  more common  `shooting stars' (i.e., papers that are highly cited initially but die quickly), can be explained by tunning a few parameters of mechanistic models of the citation dynamics \cite{Mingers_2008,Min_2018,Wang_2013}.  A seemingly  natural mechanistic model  to describe the spread of ideas in the academia is the SIR epidemic model  \cite{Bettencourt_2006}, which, however, has not yet been applied  to the analysis of the citation histories of individual papers.   

Accordingly, here we use  the SIR epidemic model to fit the  number of citations  received over a period of 15 years by 300 hit papers  published in 6 selected APS  journals. We define hit papers as the 50 most cited papers published  from 2000 to 2003 in each  of those journals. The data used in our analysis is from the American Physical Society Data Sets for Research (available upon request at \cite{APS_Dataset}), which  include only internal citations, i.e., citations to papers published in APS journals   from  papers  published  in APS  journals as well. In  Sec.\ \ref{sec:dataset}, we offer a brief overview of the APS Dataset.
 
The epidemiological parameters of the SIR epidemic  model have a direct interpretation in terms of the citation dynamics, as we explain in great detail in Sec.\ \ref{sec:model}. Following the maxim `citations beget citations'  we assume that the citations of a given paper are promoted   by  a certain number of influential papers whose bibliographies include that  paper and whose potential to influence yet-to-be-written papers  to cite it is determined by the transmission parameter $\beta$. Papers cease to be influential at a rate $\gamma$.   The ratio $R_0 = \beta/\gamma$ is very close to 1 for most of the hit papers considered here, so that the `shooting stars'  and the `sleeping beauties' citation patterns are obtained for large and small  values of $\beta$, respectively.   An  epidemiological parameter of particular interest is the  potential maximum  number of citations $S_0$ a hit paper can receive, which yields an estimate of the size of the community that could in principle cite that paper.   The   
SIR model allows the ready estimate of the total number of  citations a paper ever acquires  $\Upsilon_\infty \leq  S_0$, which can be seen  as the ultimate impact of a paper \cite{Wang_2013}. The relative ultimate impact $\Upsilon_\infty/S_0$  happens to depend  on the basic reproductive number $R_0 $ only. 

The distributions of the epidemiological parameters for each selected journal enables the ranking of the journals using  their medians, as we describe in Sec. \ref{sec:rank}. In particular,  the ranks produced by the medians of the distributions of the parameters $S_0$ and $\beta$  interchanges  the position of two journals  only in comparison with  the Impact Factor (IF) rank \cite{Garfield_2006}. In general, however, the higher the IF of a journal is, the less its relative ultimate impact.

\section{APS Dataset}\label{sec:dataset}

The APS Dataset comprises citing article pairs and article metadata of about $636000$  papers published in  $17$ journals of the American Physical Society from $1893$ to $2018$ \cite{APS_Dataset}.
The first  journal, \textit{Physical Review}, ceased publication  in $1969$, and the most recent  journal in the dataset, \textit{Physical Review Materials}, was launched  in $2017$.

\begin{figure}[t!]
\centering  
\includegraphics[width=0.45\textwidth]{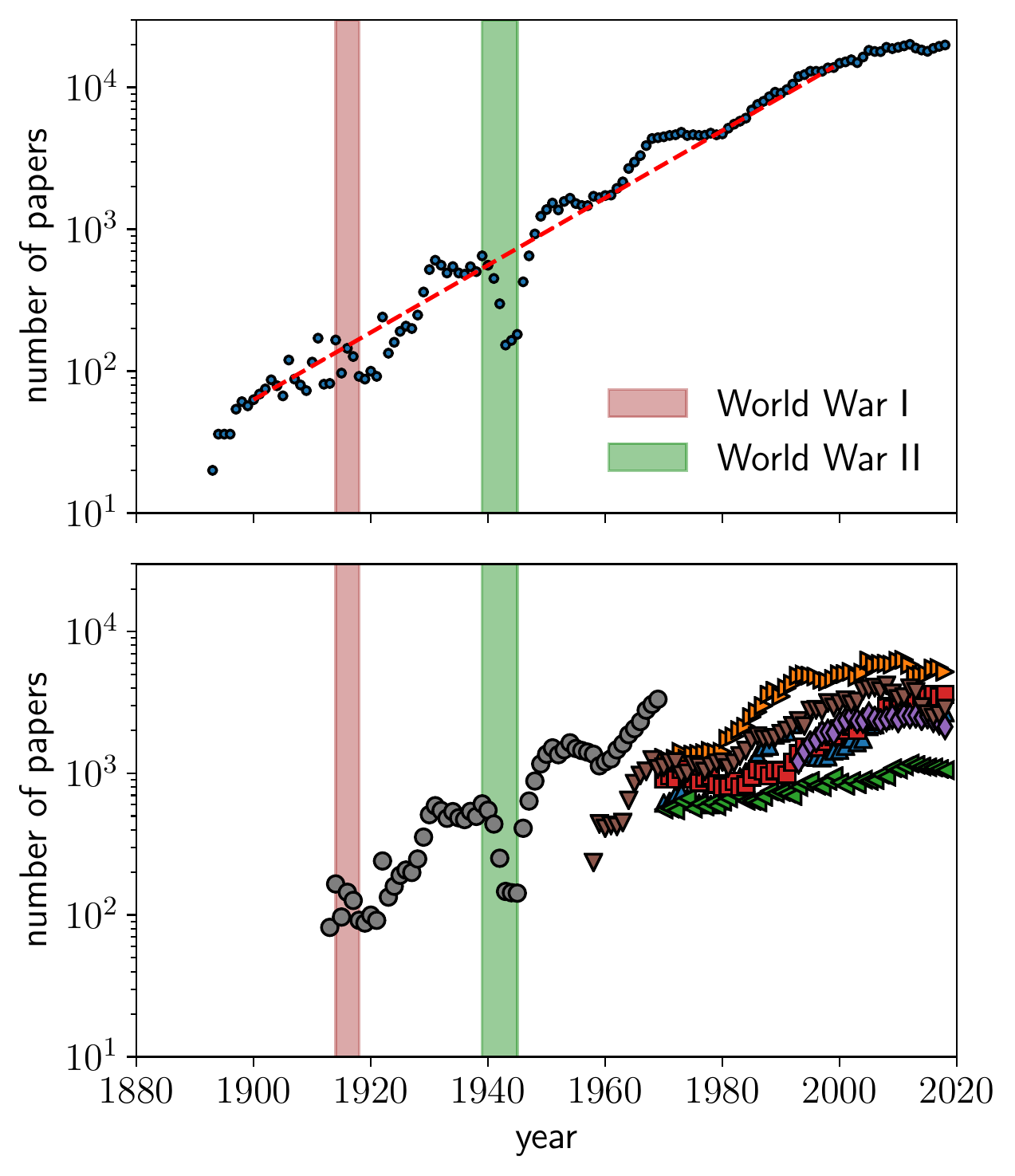}
\caption{Number of papers published in all APS journals from $1893$ to $2018$  (upper panel). The dashed straight line is  the fitting function $f(x) = a \exp \left [ (x - 1900)/b \right ]$ with  $a = 63$ and $b = 18$.  The lower panel shows the number of papers published in 
 \textit{Phys.\ Rev.\ Series II} ($\circ$),  \textit{Phys.\ Rev.\ Lett. } ($\triangledown$), \textit{Phys.\ Rev.\ A} ($\triangle$), \textit{B} ($\triangleright$), \textit{C} ($\triangleleft$), \textit{ D} ($\Box$) and  \textit{E} ($\diamond$).
}
\label{fig1_APS}
\end{figure}

With more than one century of existence, the APS journals lived through the World War I and the World War II. The total number of papers 
published per year  in the APS journals shown in the upper panel of  Fig.\ \ref{fig1_APS} reveals the  distinct  impact these two major events had on the academic
productivity of the physicists.   Whereas WWI caused no discernible changes on the number of papers published in the \textit{Physical Review}, most likely because this journal was not yet the main publication choice of  the European physicists,    WWII caused a sharp drop on the number of published papers, which reflects the worldwide disruption this event  produced in all activities unrelated to warfare.   More importantly, this panel shows  that the number of published papers grows at an exponential rate \cite{Price_1975}, which probably prompted the splitting of the \textit{Physical Review Series II} 
into \textit{Physical Review A}, \textit{B}, \textit{C}, and \textit{D} in 1970. Nevertheless, the exponential growth trend   continued for the offspring journals, leading to the current difficulty of physicists to keep pace with the advances of their own research subfields \cite{Larsen_2010,Landhuis_2016}.  The  lower panel of Fig.\ \ref{fig1_APS} shows the number of papers published in the APS journals we will consider in this paper. In addition to the journals already mentioned, the panel includes the  \textit{Physical Review Letters} that was introduced in 1958 and the 
\textit{Physical Review E} that  was launched in 1993. For completeness, we also include in the panel  the  \textit{Physical Review, Series II} which replaced the \textit{Physical Review, Series I} and was active from $1913$ to $1969$.  Those are the $7$ APS journals with the largest number of papers published since $1913$.

\begin{table} [h!]
\caption{Number of papers published from $2000$ to $2018$ in the  $6$ APS journals  used  in our citation dynamics analysis.}
\centering
\begin{tabular}{| l || c |}
\hline 
APS Journal & number of papers  \\
\hline
\textit{Phys.\ Rev.\ A}        &   5826        \\
\textit{Phys.\ Rev.\ B}        &   18028        \\
\textit{Phys.\ Rev.\ C}       &   3131         \\
\textit{Phys.\ Rev.\ D}        &   7589         \\
\textit{Phys.\ Rev.\ E}        &   8137         \\
\textit{Phys.\ Rev.\ Lett.}    &   12163        \\
\hline
\end{tabular}
\label{table1}
\end{table}
 
  Here we  focus on the $54874$ papers published in  \textit{Phys.\ Rev.\ Lett.}, \textit{Phys.\ Rev.\ A},  \textit{B}, \textit{ C}, \textit{D}, and  \textit{E} from $2000$ to $2018$. Table \ref{table1}  shows the number of papers published in each of these journals in that time window. 
For each journal we pick  the $50$  papers  published from $2000$ to $2003$ that received the highest number of citations up to $15$ years (180 months) after their publication dates. Those papers are named hit papers and next we will show how to model their citation patterns using an epidemiological model.  We refer the reader to Ref.\ \cite{Redner_2005} for the analysis of the citation statistics of Physical Review from 1893 through 2003.

\section{Epidemiological model}\label{sec:model}

\begin{figure}[!ht]
\centering
\includegraphics[width=0.45\textwidth]{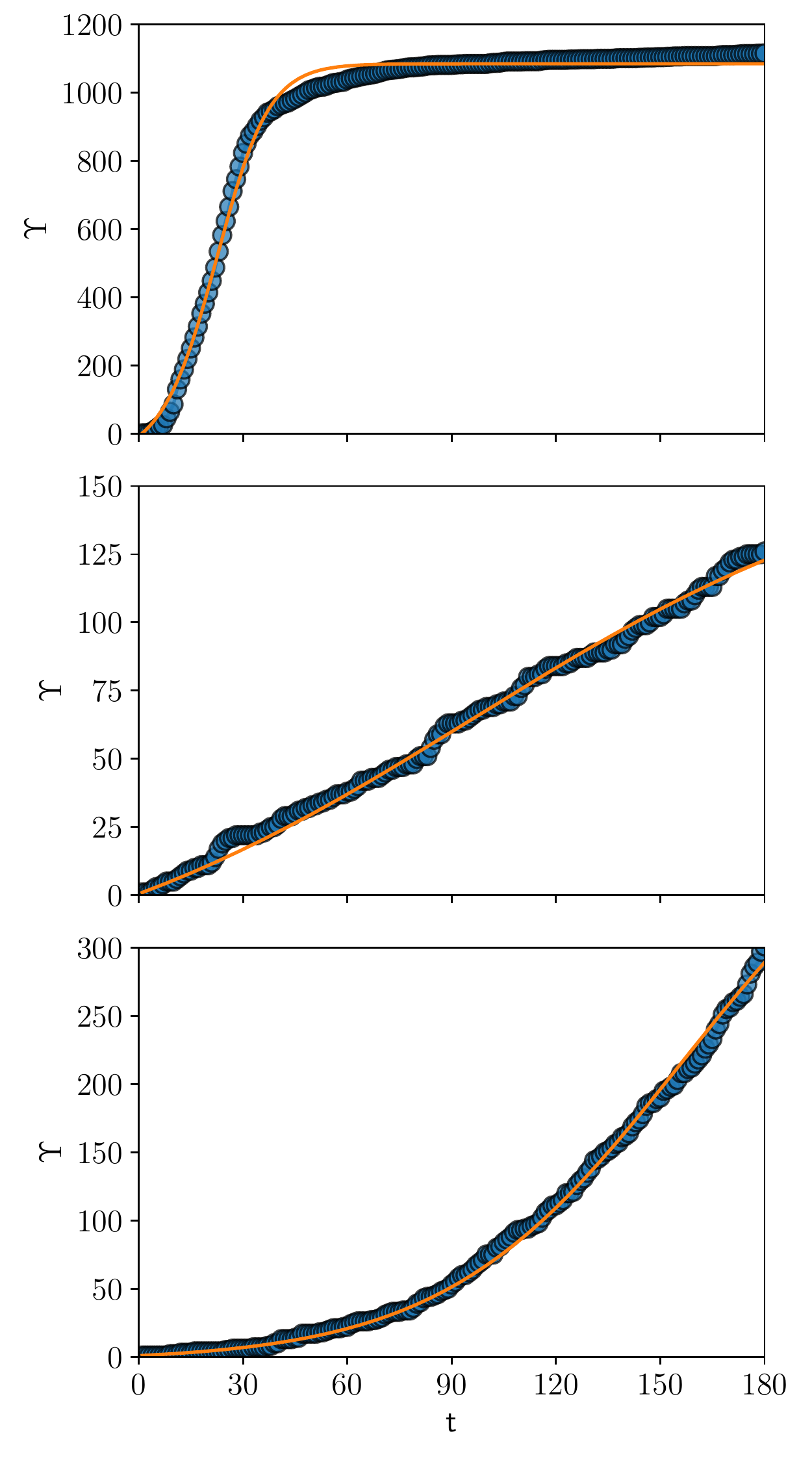}  	  
\caption{Cumulative   number of citations $\Upsilon$ of three representative hit papers as function of the time $t$, measured in  months,  after their publication dates.  The symbols are the citation data extracted from the APS dataset  and  the solid curves are the fittings with the SIR model. The epidemiological parameters $[S_0, \beta, \gamma]$ are   $[42000, 9.36, 9.25]$ (upper panel), $[3150, 0.48, 0.47]$ (middle panel) and  $[1050, 0.13, 0.10]$ (lower panel).  
 }
\label{fig:epi1}
\end{figure}

We  characterize the citation histories of the hit papers  by their cumulative number of citations received in the  period of 180 months from their  publication dates.  Figure \ref{fig:epi1} illustrates this quantity for three representative  citation histories in the time window considered. In particular, the number of citations of the paper  shown in the upper panel \cite{PhysRevD.66.010001}  exhibits a very rapid increase  followed by stabilization  within about 30 months after its publication.  This is the editors' dream paper  for it has the perfect timing to boost the IF of a journal. The citation record of the paper  shown in the middle panel  \cite{PhysRevA.65.032323} exhibits a steady and consistent growth of little less than one citation per month. Perhaps the most interesting citation pattern is that of the paper  exhibited in the lower panel \cite{PhysRevB.65.035109},  which displays 
 a latent period  followed by a steady speed up  of the number of citations. This panel  illustrates the   `sleeping beauties' citation pattern  that makes the prediction of the impact of a research using short-time  information (say, a 24 months window) a somewhat shortsighted enterprise \cite{van2004sleeping,ke2015defining}.

The  main  assumption behind our approach to model the citation histories illustrated in Fig.\  \ref{fig:epi1} is that citations beget more citations,  so that the citation dynamics could be modeled as the spread of an infectious disease. This means that a  particular hit  paper comes to  the knowledge of prospective citing authors through the reading of   papers that cite the hit paper.   Here we model the citation dynamics of a hit paper using the popular SIR model  \cite{kermack1927contribution,Murray1993} where the susceptible (S), infected (I) and removed  (R) classes must be properly reinterpreted within  the citation dynamics context. 

In particular, once a  hit paper   is published we assume that there is a   maximum  number of citations it can receive, which we denote by $S_0$. This is the number of papers in an abstract population of  papers not yet written that are susceptible to cite the hit paper.  This number can only decrease with time and we denote by  $S(t) \leq S_0$ the  number of  citations  the hit paper can still receive after time $t$  from its publication date. Of course, 
\begin{equation}\label{Y0}
\Upsilon(t) = S_0 - S(t)
\end{equation}
 is the  measurable total number of citations the hit paper received until time $t$, which is shown in Fig.\ \ref{fig:epi1} for three selected hit papers.
 In principle, $S_0$ could be estimated if we knew the size of the community that works on the subject addressed by the hit paper, the mean number of papers published per month by  researchers in that community and  the average number of references those papers contain.

A simple way to model the  decrease of the number of susceptible papers with time is  through the contact process
\begin{equation}\label{S}
\frac{dS}{dt}  =   - \beta S \frac{I}{N} ,
\end{equation}
where $I=I(t)$ is the number of papers that have cited the hit paper  before or at  time $t$ and that  can still influence susceptible papers  to cite that paper. In other words, $I(t)$ is the number of influential (or infective) papers.   Although the hit paper does not cite itself, we will assume that it contributes to $I(0) = I_0$.  We note that  the hit paper may not contribute to  $I(t)$, i.e., it may not be  influential any more at time $t>0$ so that its new citations are prompted by third-party papers.  (Just think of the many papers citing classic books whose authors never read those books.)   The  coefficient $\beta$ in Eq.\ (\ref{S})  is  a measure of the persuasion power of the influential papers, i.e., it is a measure of   the  likelihood an author will cite the hit paper because that author read a paper that cites the hit paper. Because $\beta$  is a per capita transmission rate we have  introduced the constant  factor $N=S_0 + I_0$ in Eq.\ (\ref{S}) to guarantee that its value  is on the order of 1 regardless of the value of $S_0$,  and that  this equation is dimensionally correct. (The unit of $S$ and  $I$ is papers.)

The equation for the number of influential papers 
\begin{equation}\label{I}
\frac{dI}{dt}  =   \beta S \frac{I}{N}  - \gamma I  
\end{equation}
makes plain the fair assumption  that influential papers cease to be influential at a rate  $\gamma$ and  move into the removed class. Papers in the removed class play  no role in the citation dynamics  and their number is given by  
\begin{equation}\label{R}
\frac{dR}{dt}  =    \gamma I . 
\end{equation}
Since $d(S+I+R)/dt = 0$ we have  $S(t) + I(t) + R(t)= S_0 + I_0 = N$,  because the removed class is empty at $t=0$. Moreover,  since the number of citations are  reported on a monthly basis, we use the month as our  time unit  so that $\beta$ and $\gamma$ have unit 1/month.

We note that our epidemiological approach builds  on an   assumption different from the vastly popular cumulative advantage or preferential attachment assumption,  in which the probability
that a publication is cited is an increasing function of its
current total number of citations \cite{Redner_2005, Merton_1973}. In our case, this probability is a function of  a (variable) fraction of the total number of citations, namely, it is a function of the number of influential papers.

 Perhaps the  most interesting quantity in the citation dynamics context is   $ \Upsilon_\infty = \lim_{t \to \infty} \left [ S_0 - S(t) \right ]  = S_0 - S_\infty$, which gives the total number of citations a hit  paper acquires during
its lifetime, i.e., its ultimate impact. Of course, $ \Upsilon_\infty$ cannot be measured but can be easily inferred  using our epidemiological approach. In fact, it is given by the positive root of the transcendental equation \cite{Murray1993}
\begin{equation}\label{Y1}
\Upsilon_\infty = S_0 \left [ 1 - \exp ( -  \frac{ \Upsilon_\infty +I_0}{N \rho} ) \right ]
\end{equation}
where $\rho = \gamma/\beta$. Hence $\Upsilon_\infty = S_0$ only in the limit $\rho \to 0$. For finite $\rho$,  a hit paper will receive only the fraction $\upsilon = \Upsilon_\infty/S_0$ of the  potential citations it could receive.  For instance, for the papers analyzed in Fig.\  \ref{fig:epi1}, the estimated total  number of citations they  will receive is $\Upsilon_\infty = 1060$ (upper panel), $166$ (middle panel) and $446$ (lower panel).

Assuming that $I_0 \ll N \approx S_0$ we rewrite Eq.\ (\ref{Y1}) as   
\begin{equation}\label{Y2}
\upsilon = 1 - \exp ( -  \frac{ \upsilon}{\rho} ),
\end{equation}
which has a nonzero solution provided that $R_0 = 1/\rho > 1$. Here $R_0$ is the basic reproductive  number that ultimately determines the overall  impact of a hit paper on the abstract population of susceptible papers. Since our focus are on  hit papers only, we have $R_0 > 1$ necessarily. In fact, with very few exceptions, the values of $R_0$ of the hit papers were all very close to 1.

The SIR model has only three adjustable parameters, namely, $S_0$, $\beta$ and $\gamma$ and our goal is to estimate these parameters  by fitting   Eq.\ (\ref{Y0}) to the cumulative number of citations extracted from the APS dataset.  
The quality of the  fitting  can be appreciated in Fig.\  \ref{fig:epi1}. Incidentally, the paper considered in the upper panel  of this figure has the largest value of $\beta$
among the 300 hit papers considered in our study.


\begin{figure}[t]
\centering  
\includegraphics[width=0.45\textwidth]{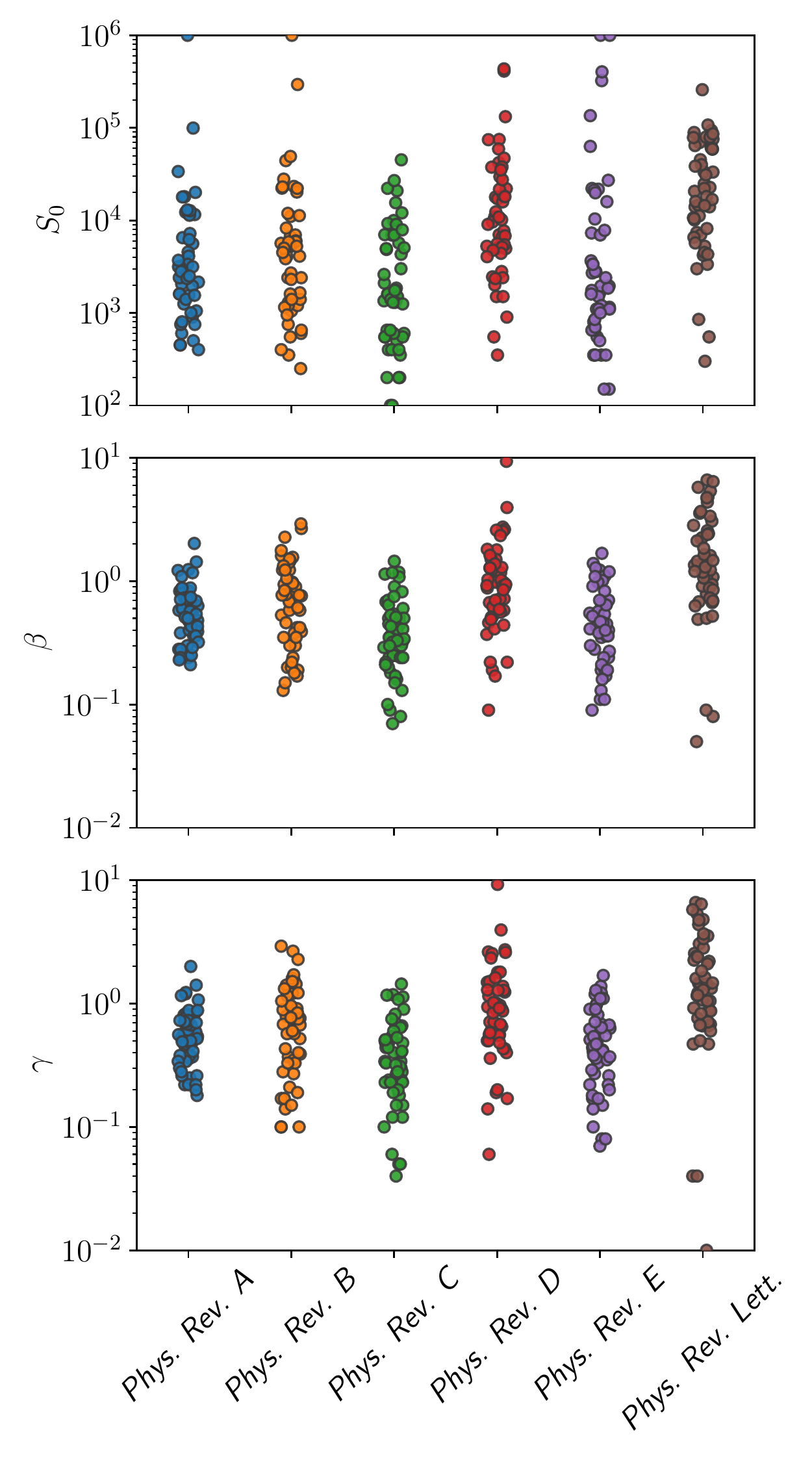}
\caption{Epidemiological parameters $S_0$, $\beta$ and $\gamma$  that best fit the theoretical estimate of  $\Upsilon (t)$  to the empirical cumulative citation numbers. Each symbol  corresponds to a particular hit paper  published in the indicated APS journal.  The unit of $\beta$ and $\gamma$ is 1/month and the unit of $S_0$ is papers.
  }
\label{fig:SIR}
\end{figure}

\section{Epidemiological journal ranking}\label{sec:rank}

The results of our fitting procedure are summarized  in Fig.\ \ref{fig:SIR}, which shows the epidemiological parameters $S_0$,  $\beta$  and $\gamma$ that best fit the theoretical estimate of $\Upsilon (t)$  to the empirical cumulative number of citations.
Each symbol in a panel corresponds to the estimated epidemiological parameter of a particular hit paper.  Because of the considerable spread of the values of the estimates,  which is particularly pronounced for $S_0$, it is convenient to summarize  the parameter distributions by their medians, $\tilde{S_0}$, $\tilde{\beta}$ and $\tilde{\gamma}$, which are shown in Table \ref{table2} together with the medians of the basic reproductive number $\tilde{R}_0$ and of the ultimate fraction of the potential  citations received $\tilde{\upsilon}$.  The order of the journals listed  in this table is determined by the value of  $\tilde{S_0}$.

\begin{table} [h!]
\caption{Medians of the potential number of citations  ($\tilde{S}_0$),  the per capita  transmission rate ($\tilde{\beta}$), the removal rate  ($\tilde{\gamma}$),  the basic reproductive number ($\tilde{R}_0$) and the fraction of the potential  citations received ($\tilde{\upsilon}$) for the selected  APS journals. }
\centering
\begin{tabular}{| l || c | c | c | c | c |c |}
\hline 
APS Journal & $\tilde{S}_0$  & $\tilde{\beta}$ &  $\tilde{\gamma}$ & $\tilde{R}_0$ &  $\tilde{\upsilon}$  \\
\hline
\textit{Phys.\ Rev.\ Lett.} & 19350 & 1.400 & 1.385 & 1.008 & 0.021 \\
\textit{Phys.\ Rev.\ D} & 9325 & 0.915 & 0.910 & 1.012 & 0.031 \\ 
\textit{Phys.\ Rev.\ B} & 4750 & 0.730 & 0.715 & 1.026 & 0.059 \\ 
\textit{Phys.\ Rev.\ A} & 2900 & 0.570 & 0.550 & 1.031 & 0.072 \\ 
\textit{Phys.\ Rev.\ E} & 1900 & 0.455 & 0.445 & 1.028 & 0.070 \\
\textit{Phys.\ Rev.\ C} & 1600 & 0.355 & 0.340 & 1.037 & 0.084 \\ 
\hline
\end{tabular}
\label{table2}
\end{table}

\begin{figure}[t]
\centering  
\includegraphics[width=0.45\textwidth]{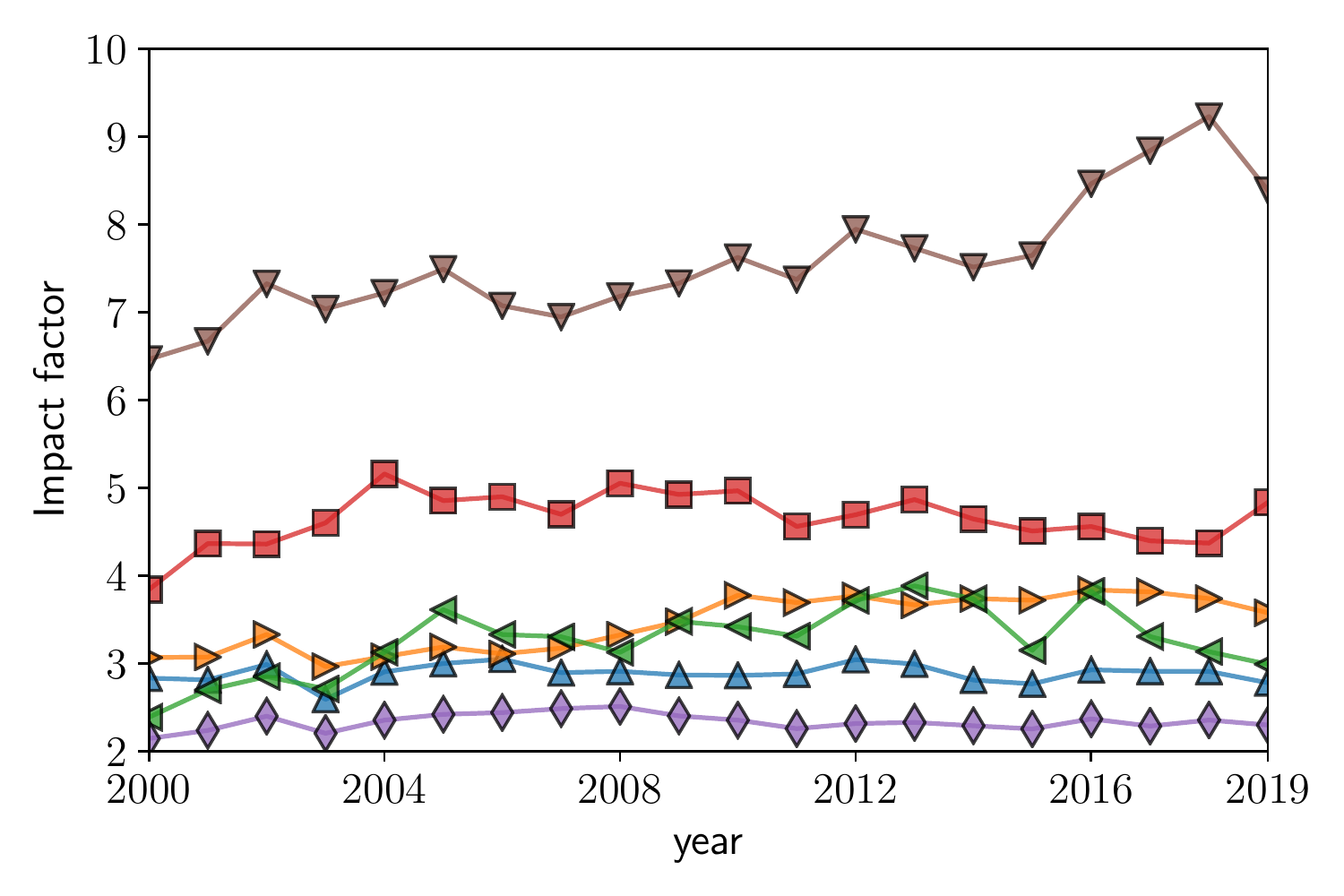}
\caption{Evolution of the Impact Factor  from 2000 to 2019 of the  journals  \textit{Phys.\ Rev.\ Lett. } ($\triangledown$), \textit{Phys.\ Rev.\ A} ($\triangle$), \textit{B} ($\triangleright$), \textit{C} ($\triangleleft$), \textit{ D} ($\Box$) and  \textit{E} ($\diamond$). The hit papers considered in this paper were published between 2000 and 2003.  }
\label{fig:IF}
\end{figure}

Remarkably, Table \ref{table2} shows that $\tilde{S_0}$, $\tilde{\beta}$  and $\tilde{\gamma}$ are good predictors of the rank of  the selected APS journals according to the IF metric \cite{Garfield_2006}.  In fact, if we consider that the hit papers were published between 2000 and 2003, the epidemiological rank  offered in this table interchanges the positions of \textit{Phys.\ Rev.\ C} and  \textit{Phys.\ Rev.\ E} only as compared with  the IF rank (see Fig.\ \ref{fig:IF}).  The good agreement between these ranks is not very surprising in the sense that it is well known that the size of the community, which in  our approach is measured by $S_0$, correlates well with the IF
\cite{Althouse_09}. It is important to note, however,  that $S_0$ does not correlate well with the number of papers published in a journal during the period of analysis shown in Table \ref{table1}.
 In addition and perhaps more importantly, because the IF is measured in  a 24 months window, it correlates well with the transmission rate $\beta$, since large values of this parameter result in  many  citations  in a short time provided there are plenty of  susceptible papers  (see upper panel of Fig. \ref{fig:epi1}). These  findings validate our theoretical approach to model the citation dynamics  as well as the procedure we used to estimate the epidemiological parameters. 
 
 The positive correlation between the IF metric and  $\tilde{\gamma}$ is more intriguing and, perhaps, illuminating.  We recall that a high value of $\gamma$ implies a quick obsolescence of the influential papers, i.e.,  those papers are influential for a short  period of time only. This means that the epidemiological approach predicts that papers that cite hit papers published in high impact factor journals are not likely to be  very impactful themselves.  This scenario is evocative of the  many application papers that use a novel analysis method presented in a hit paper.

The surprising finding revealed in Table \ref{table2} is the negative correlation between    $\tilde{\beta}$ (and hence the IF metric) and $\tilde{R}_0$ or $\tilde{\upsilon}$. We recall that  $R_0$ and $\upsilon$  are related through  Eq.\ (\ref{Y2}). Actually the true relevance of $R_0$  can be appreciated only through  its link to $\upsilon$,  which reveals that $R_0 > 1$ does not imply that an infective agent (a hit paper in our case) will take over the entire susceptible population. Interestingly,  our epidemiological analysis indicates that, when the size of the susceptible community $S_0$ is taken into account,  hit papers published in high impact journals   actually have a smaller (relative) number of citations than hit papers published in   low impact journals. 
We note that  $\tilde{R}_0$ and  $\tilde{\upsilon}$ are not related by  Eq.\ (\ref{Y2}): these quantities are estimated from the distributions of $R_0$ and $\upsilon$ of the hit papers for each journal.

\section{Conclusion}\label{sec:conc}

The  literature   already offers several  mechanistic models for the citation dynamics of individual papers. Some  of them build on the similarity between the S-shaped curves of the cumulative number of  citations  and the curves that describe the  diffusion of innovations  to argue that the same mechanisms that drive the adoption  of a new product, viz.,  innovation and imitation  \cite{Rogers_2010,Bass_1969}, may explain the citation process as well \cite{Mingers_2008,Min_2018}. 
However, the likely most  successful mechanistic model of citation dynamics builds on assumptions proper to this dynamics, viz., preferential attachment,  fitness and  aging   \cite{Wang_2013}.  As already pointed out, preferential attachment or  cumulative advantage means that the probability
that a publication is cited is an increasing function of its current number of citations \cite{Price_1975,Merton_1973,Redner_2005}. Fitness expresses the notion that papers differ with respect to the perceived novelty and importance of their contents \cite{Foster_2015,Li_2019} and aging captures the fact  that the perceived novelty and importance of a paper  eventually  fade out \cite{Eom_2011}. There are, of course, many intangible factors behind an author's decision to cite a paper, such as the  reputation of its authors and the  journal where it was published, that can be identified in a citation network  analysis 
but cannot be implemented in a mechanistic model  \cite{Dong_2016}.

Here we take a different approach that  is inspired by the attempt to describe the spread  of Feynman diagrams through the theoretical physics communities of different countries using models of epidemics \cite{Bettencourt_2006}. In particular,  we fit the citation history of 300 hit papers from 6 selected APS journals using a SIR epidemic  model. The advantage of this approach is that the epidemiological parameters have a direct interpretation in terms of the citation dynamics. For instance, a paper's relative long term impact   (i.e.,  the total number of citations a paper will ever acquire) is easily derived within the epidemiological framework  (see Eq.\ (\ref{Y2}))  and it is a function of the basic   reproductive number $R_0 = \beta/\gamma$ only.  This result is similar to the  ultimate impact of a paper derived in Ref.\  \cite{Wang_2013}, which happens to depend only on the relative fitness of the paper.  In fact, recalling that in the  infectious disease context $R_0$ is the average number of people infected from one person and that in the evolutionary context the fitness of an organism is measured by the number of offspring per generation it produces,  it is fair to think of $R_0$ as the fitness of the paper,  so the two distinct approaches reach similar conclusions.  However, the epidemiological approach does not make the preferential attachment assumption, since it assumes  that  the probability  that a publication is cited at a certain time is a function of the number of influential papers  at that time, which is a time-dependent fraction of  the current number of citations.  

The distributions of the values  of the epidemiological parameters  that describe the citation histories of the 50 hit papers   for each one of the 6 APS journals considered  allow us to characterize  those journals   and  define an epidemiological rank. It turns out  that there is a  good,  though not perfect, correlation between the rank obtained using the transmission rate $\beta$ or the potential maximum number of citations $S_0$ a paper acquires  and the IF rank. Surprisingly, this rank correlates negatively with  the rank obtained using  the basic   reproductive number $R_0$, which implies that hit papers published in high impact factor journals have less relative long-term impact ($\upsilon = \Upsilon_\infty/S_0$)  than hit papers published in low impact factor journals, although their absolute  long-term impact  ($\Upsilon_\infty$) is much greater. The fact that the IF rank is obtained using citations from papers published  in all journals indexed at the Web of Science 
may explain the discrepancies  with the epidemiological ranks that use data of  APS journals only, particularly in research areas such as Nuclear Physics where there are many traditional journals owned by other publishers.

In summary,  the SIR epidemic model  proved  very valuable to fit  the citation histories of hit papers and, in addition,  offered unexpected  insights on the citation dynamics.  The good correlation between the IF rank and the epidemiological ranks 
suggests that this simple epidemic model succeeded in picking out  the essential elements behind  the citation dynamics.

\section*{Acknowledgments}
We thank the American Physical Society for letting us use their citation database.
The research of JFF was  supported in part 
 by Grant No.\  2020/03041-3, Fun\-da\-\c{c}\~ao de Amparo \`a Pesquisa do Estado de S\~ao Paulo 
(FAPESP) and  by Grant No.\ 305058/2017-7, Conselho Nacional de Desenvolvimento 
Cient\'{\i}\-fi\-co e Tecnol\'ogico (CNPq).
SMR was supported by the Coordena\c{c}\~ao de Aperfei\c{c}oamento de Pessoal de
N\'{\i}vel Superior - Brasil (CAPES) - Finance Code 001.


\begin{thebibliography}{99}

%
%

\bibitem{Price_1975}
D. J. S.  Price,
\textit{Science since Babylon}
 (Yale University Press, New Haven,  1975) . 
 
 \bibitem{Merton_1973}
 R. Merton, 
 \textit{The Sociology of Science}
 (University of Chicago Press, Chicago, 1973).

\bibitem{Radicchi_2008}
F. Radicchi, S. Fortunato and C. Castellano,
Proc. Natl. Acad. Sci. USA \textbf{105}, 17268 (2008).

\bibitem{Radicchi_2009}
 F. Radicchi, S. Fortunato, B. Markines, and A. Vespignani,
Phys. Rev. E  \textbf{80}, 056103 (2009).

\bibitem{Uzzi_2013}
B. Uzzi, S. Mukherjee, M. Stringer and B. Jones, 
Science \textbf{342}, 468 (2013).

\bibitem{Ioannidis_2014}
J. Ioannidis, K. W. Boyack, H. Small,  A. A. Sorensen, 
and R. Klavans, 
Nature  \textbf{514}, 561 (2014).

\bibitem{Mingers_2015}
J. Mingers and L. Leydesdorff,
Eur. J. Oper. Res. \textbf{246}, 1 (2015).

\bibitem{Redner_1998}
S. Redner,
Eur. Phys. J. B \textbf{4}, 131 (1988).

 \bibitem{Redner_2005}
 S. Redner,
 Phys. Today \textbf{58}, 49  (2005).

\bibitem{Wallace_2009}
M. L. Wallace, V. Larivi\`ere and Y. Gingras,
J. Informetr. \textbf{3}, 296 (2009).

\bibitem{van2004sleeping}
A. F. J. van Raan, 
Scientometrics \textbf{59}, 467 (2004).

\bibitem{ke2015defining}
Q. Ke, E. Ferrara, F. Radicchi and A. Flammini, 
Proc. Natl. Acad. Sci. USA \textbf{112}, 7426 (2015).

\bibitem{Mingers_2008}
J.  Mingers,
J. Oper. Res. Soc.  \textbf{59},  1013 (2008).

\bibitem{Min_2018}
C. Min, Y.  Ding,  J.  Li, Y.  Bu,  L.  Pei,  and J. Sun, 
J. Assoc. Inf. Sci. Technol. \textbf{69}, 1271 (2018).

\bibitem{Wang_2013}
D. Wang,  C. Song,  A.-L. Barab\'asi,
Science   \textbf{342}, 127 (2013).

\bibitem{Bettencourt_2006}
L. M. Bettencourt, A. Cintr\'on-Arias, D. I. Kaiser and C. Castillo-Ch\'avez, 
Physica A \textbf{364}, 513 (2006).


\bibitem{APS_Dataset}
https://journals.aps.org/datasets.


\bibitem{Garfield_2006}
E. Garfield,
JAMA  \textbf{295}, 90 (2006).




%
%



 

 
\bibitem{Larsen_2010}
P. Larsen and M. Von Ins, 
Scientometrics \textbf{84}, 575 (2010).

\bibitem{Landhuis_2016}
E. Landhuis, 
Nature \textbf{535}, 457 (2016).
 

%
%
\bibitem{PhysRevD.66.010001}
K. Hagiwara et al.,
Phys. Rev. D  \textbf{66}, 010001 (2002).
 
 \bibitem{PhysRevA.65.032323}
 M. S. Kim,  W. Son,  V. Bu\ifmmode \check{z}\else \v{z}\fi{}ek  and P. L. Knight, 
 Phys. Rev. A   \textbf{65}, 032323 (2002).
 
 \bibitem{PhysRevB.65.035109}
I.  Souza,  N. Marzari and D. Vanderbilt, 
 Phys. Rev. B  \textbf{65}, 035109 (2001).
 
 

\bibitem{kermack1927contribution}
W. O. Kermack and A. G. McKendrick,
Proc. R. Soc. A  \textbf{115}, 700 (1927).

\bibitem{Murray1993}
J. D. Murray, 
\textit{Mathematical Biology I: An Introduction} 
(Springer,  New York, 1993).

\bibitem{Althouse_09}
B. M. Althouse,  J. D. West,  C. T. Bergstrom  and T. Bergstrom,
J. Assoc. Inf. Sci. Technol. \textbf{60}, 27 (2009).

%
%

\bibitem{Rogers_2010}
E. M. Rogers, \textit{Diffusion of Innovations} 
(Simon and Schuster,  New York, 2010).

\bibitem{Bass_1969}
F. M. Bass,   
Manag. Sci.  \textbf{15}, 215 (1969).



\bibitem{Foster_2015}
J. G. Foster, A. Rzhetsky, and J. A. Evans, 
Am. Sociol. Rev. \textbf{80}, 875 (2015).

\bibitem{Li_2019}
J. Li, Y. Yin, S. Fortunato and D. Wang, 
Nat. Rev. Phys. \textbf{1}, 301 (2019).


\bibitem{Eom_2011}
Y.-H. Eom and S.  Fortunato,
PLoS ONE \textbf{6}, e24926 (2011).

\bibitem{Dong_2016}
Y. Dong, R. A. Johnson, and N. V. Chawla, 
IEEE Trans. Big Data \textbf{2}, 18 (2016).









\end{thebibliography}
\end{document}